\documentclass[journal=apchd5,manuscript=article]{achemso}

\usepackage{achemso}

\usepackage{lmodern} 
\usepackage{microtype}

\usepackage[utf8]{inputenc}
\usepackage{amsmath}
\usepackage{amssymb}
\usepackage{caption}
\usepackage{subcaption}
\usepackage[capitalise]{cleveref}
\usepackage{float}

\DeclareMathOperator{\argmin}{arg\,min}
\DeclareMathOperator{\sgn}{sgn}
\DeclareMathOperator{\support}{support}
\usepackage{amsmath}
\usepackage{caption}
\usepackage{subcaption}
\usepackage[capitalise]{cleveref}
\usepackage{setspace}
\usepackage{url}

\usepackage{float}

\def\figures{}

\author{Gaurav Arya}
\email{aryag@mit.edu}

\affiliation[MIT Math] {Department of Mathematics, Massachusetts Institute of Technology}
\author{William F. Li}
\affiliation[MIT RLE] {Research Laboratory of Electronics, Massachusetts Institute of Technology}

\author{Charles Roques-Carmes} 
\affiliation[Stanford]
{E. L. Ginzton Laboratory, Stanford University}
\alsoaffiliation[MIT RLE] {Research Laboratory of Electronics, Massachusetts Institute of Technology}
\author{Marin Solja\v{c}i\'{c}}
\affiliation[MIT RLE] {Research Laboratory of Electronics, Massachusetts Institute of Technology}
\alsoaffiliation[MIT Physics] {Department of Physics, Massachusetts Institute of Technology}
\author{Steven G. Johnson}
\affiliation[MIT Math] {Department of Mathematics, Massachusetts Institute of Technology}

\author{Zin Lin}
\affiliation[Vtech] {Bradley Department of Electrical and Computer Engineering, Virginia Tech}

\title{End-to-End Optimization of Metasurfaces for Imaging with Compressed Sensing}

\begin{document}




\abstract{
We present a framework for the end-to-end optimization of metasurface imaging systems that reconstruct targets using compressed sensing,
a technique for solving underdetermined imaging problems when the target object exhibits sparsity ({e.g.} the object can be described by a small number of nonzero values, but the positions of these values are unknown). We nest an iterative, unapproximated compressed sensing reconstruction algorithm into our end-to-end optimization pipeline, resulting in an interpretable, data-efficient method for maximally leveraging metaoptics to exploit object sparsity. We apply our framework to super-resolution imaging and high-resolution depth imaging with a phase-change material. In both situations, our end-to-end framework {effectively optimizes} metasurface structures for compressed sensing recovery, automatically balancing a number of complicated design considerations to select
{an} imaging measurement matrix from a complex, physically-constrained manifold with millions of dimensions. The optimized metasurface imaging systems are robust to noise, significantly improving over random scattering surfaces and approaching the ideal compressed sensing performance of a Gaussian matrix, showing how a \emph{physical} metasurface system can demonstrably approach the \emph{mathematical} limits of compressed sensing.
}



\maketitle

\section{Introduction}

In this article, we combine a subwavelength flat-optics (``metasurface'') physical platform~\cite{khorasaninejad2016metalenses, lalanne1998blazed, del2020learned, bomzon2001pancharatnam} with a compressed sensing (CS) reconstruction algorithm~\cite{4472240} to end-to-end optimize the performance of underdetermined imaging systems. 
Underdetermined imaging systems, where the imaging sensor has fewer pixels than the target object,
are becoming increasingly important as researchers attempt to extract more information from objects while keeping the imaging systems compact, inexpensive, and fast. 
CS is a powerful, principled way of solving underdetermined imaging problems
using a \emph{sparsity} prior, 
where the object (or some transformation of it) is known to have a large number of nearly zero values~\cite{4472247,willett2011compressed,Boominathan:22}. However, in CS problems, the parameters of the physical platform (which determine the ``measurement matrix'' of the system) have traditionally been assumed to be fixed.
\ifdefined\figures
\begin{figure}
    \centering
    \includegraphics[width=1\textwidth]{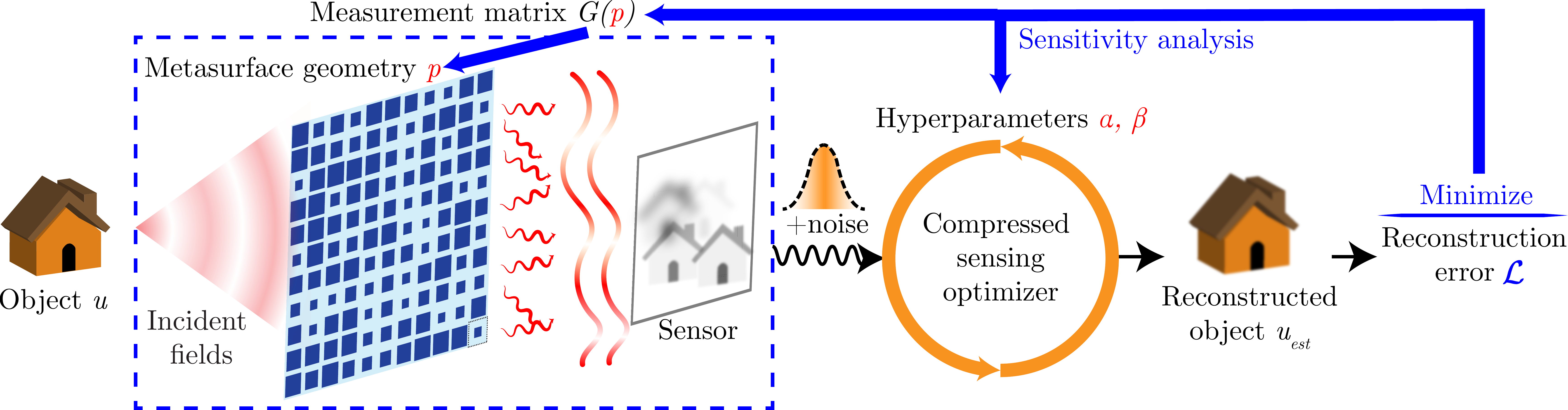}
    \caption{Bilevel optimization problem for a metasurface imaging system in an underdetermined and potentially multichannel imaging setting. A compressed sensing minimization problem is \emph{nested} within the optimization pipeline. We find the optimal measurement matrix that is physically realizable by our metasurface optics, along with the best corresponding reconstruction hyperparameters, through end-to-end optimization of the reconstruction error.}
    \label{fig:schematic}
\end{figure}
\else 
\refstepcounter{figure}\label{fig:schematic}
\fi
In this work, we show how to ``end-to-end'' optimize the metasurface imaging system to find the \emph{physically realizable} measurement matrix that minimizes the CS reconstruction error.
Thus, we propose a framework (\cref{fig:schematic})
for answering the following question: how can we maximally leverage metaoptics to exploit object sparsity?
We embed an unapproximated, iterative $\ell_1$-regularized CS algorithm into our end-to-end pipeline, performing an adjoint sensitivity analysis of the Karush-Kuhn-Tucker~(KKT) optimality conditions of the reconstructed CS solution for efficient backpropagation. 
Like any classical CS algorithm, the reconstruction process is interpretable, data-efficient, and generalizable to data outside the training distribution:
the \emph{only} assumption made is sparsity, and no additional data is needed to formulate and solve the optimization problem.
This distinguishes our approach from previous end-to-end metasurface~\cite{lin2021end, Tseng2021, lin2021end0, del2020learned} and diffractive-optics~\cite{sitzmann2018end,baek2021singleshot,markley2021physicsbased, sun2020end, tseng2021differentiable, chang2019deep} optimization that considered overdetermined problems where CS was not necessary and/or took a data-intensive deep-learning approach to image reconstruction.
  
Our end-to-end optimization forms composite lens-like metasurface configurations that dramatically improve upon random scattering surfaces. For example, for pixel-limited imaging of sparse objects with a small sensor, we discover computational metasurface designs that approach the theoretically optimal performance of physically unrealizable Gaussian measurement matrices. 
Importantly, metasurfaces exploit full-wave electromagnetic physics---such as multiple scattering, strong wavelength/polarization dispersion, and resonances~\cite{Mait:20, Engelberg2020}. Even greater inference capabilities can potentially be unlocked by phase-change materials (PCMs)~\cite{simpson2020phase,shalaginov2021reconfigurable,he2019tunable}, which allow one to collect more data with a single device.
We show that our end-to-end framework can maximize complex electromagnetic interactions in PCMs to tackle severely underdetermined multichannel imaging problems 
such as high-resolution 3D imaging. Our work represents a new paradigm for enabling \emph{maximal symbiosis} between metaoptics and CS. Our framework is also immediately applicable to diffractive optics, which are simply a special limiting case of metasurfaces (when the surface is locally uniform on the wavelength scale). 
These results open up new possibilities for high-resolution depth, spectral, and polarization imaging and beyond. 
For example, in subsequent work we have applied our end-to-end CS framework to design 3D volumetric nonparaxial metasurfaces for detecting angles and wavelengths of sparse incoming beams, yielding order-of-magnitude improvements in resolution $\times$ size~\cite{li2023transcending}.

Our metasurface physics platform is an ultrathin nanostructured interface composed of an aperiodic array of subwavelength nanopillars, similar to structures previously fabricated for various metaoptics applications~\cite{khorasaninejad2016metalenses, mueller2017metasurface, lin2021end, khorasaninejad2016polarization, li2021inverse, chen2018broadband}. In contrast to diffractive optical elements, which can be modeled by scalar diffraction theory~\cite{born2013principles}, metasurfaces consist of nanoscale elements which require the simulation of ``full-wave'' Maxwell's equations. This results in enhanced dispersion, which can be used to realize multifunctional devices~\cite{chen2018broadband, rubin2019matrix, Roques-Carmes2022}. Furthermore, metasurfaces can be fabricated over large scales with chip-scale fabrication techniques in various mature (e.g., titania~\cite{devlin2016broadband} and silicon~\cite{arbabi2015dielectric}) and emerging (e.g., PCMs ~\cite{simpson2020phase,shalaginov2021reconfigurable,he2019tunable}) material platforms.
The large number of degrees of freedom (up to $\gtrsim 10^6$ in this article) and the rich physics at the nanoscale have spurred an emerging field of large-scale metasurface inverse design~\cite{lin2018topology, molesky2018inverse, Pestourie:18}. 
However, most metaoptics designs have been confined to optics-only applications, such as lensing, beam-steering, and holography \cite{chen2016review}. Approaches combining metaoptics to a computational back-end have also been explored in the context of depth~\cite{colburn2020metasurface, yang2023monocular} and thermal imaging~\cite{saragadam2024foveated, huang2024broadband}. In those applications, the desired image is prescribed \emph{a~priori}. In contrast, recent work has introduced \emph{end-to-end} metaoptics inverse design~\cite{lin2021end0,lin2021end,Tseng2021}, whereby metasurfaces are directly optimized to minimize the reconstruction error in overdetermined imaging settings, computationally discovering the best sensor image for subsequent reconstruction.
This article extends end-to-end metasurface optimization to the underdetermined regime, where sensor data is limited, for both single-channel (2D) imaging 
with a small sensor and multichannel (3D) high-resolution depth imaging with a phase-change material.

Compressed sensing (CS) has become increasingly prevalent in computational imaging, whereby optical components are paired with a CS reconstruction algorithm, which in some sense finds the ``sparsest'' object that produces the observed image. 
How can we optimize a CS imaging system? Despite the genericness of the imaging task (full target object reconstruction), there is no clear way of formulating a \emph{heuristic} objective for this problem: 
conventional objectives such as the condition number fail in the underdetermined case (the imaging matrix is necessarily singular), while figures of merit from CS theory such as the restricted isometry property are intractable to evaluate for large matrices~\cite{bandeira2013certifying}. Standard two-point resolution metrics also fail (as noted in previous work~\cite{Antipa:18}), with the key reason being that a perfect single-foci lens is no longer an appropriate design in the undetermined regime.
{This motivates our approach:} a key contribution is to perform end-to-end optimization with a \emph{nested} CS minimization problem that is solved in a fully iterative manner (not approximated, unrolled, or truncated), as opposed to optimizing proxies for good CS performance~\cite{Yanny2020, 7882664} or employing the approximation of a differentiable unrolled network (a fixed number of iterations)~\cite{10.5555/3104322.3104374, istanet, markley2021physicsbased}. 
We solve the full bilevel optimization problem by performing a sensitivity analysis of the KKT conditions~\cite{blondel2021efficient,amos2017optnet} associated with the CS convex-optimization problem, using an adjoint formulation to efficiently backpropagate the gradient. 
Crucially, we extend previous work that performed hyperparameter optimization of CS~\cite{Bertrand_Klopfenstein_Blondel_Vaiter_Gramfort_Salmon20} by also optimizing with respect to the metasurface parameters underlying the measurement matrix itself. 
 
\section{Method}

\subsection{Bilevel Optimization Problem}
\label{sec:bilevel}

Our pipeline consists of a physics platform, which simulates light propagation through the metasurface to a grayscale sensor, and a compressed-sensing reconstruction algorithm, which solves a convex optimization problem (Fig.~\ref{fig:schematic}). We optimize this pipeline end-to-end by minimizing the expected relative mean-square error (MSE) of the reconstructions over a distribution of training objects. Given a training object $u$ (flattened into a vector), a raw noisy image $y$ is formed by:
\begin{equation}
    y = G(p) u + \eta, ~ \eta \sim \mathcal{N}(0,\sigma),
    \label{eq:noisemodel}
\end{equation}
where $G$ is the measurement matrix of the imaging system, dependent on the metasurface geometry $p$ (see below), and $\eta$ is an additive Gaussian noise~\cite{sitzmann2018end}. 
The standard deviation $\sigma$ of the Gaussian noise $\eta$ is chosen to be a few percent of the mean image intensity, i.e.~proportional to $|Gu|_1$. 
We now write a bilevel optimization problem for our end-to-end pipeline:
\begin{gather}
    \min_{p,\alpha,
    \beta} \mathcal{L} = \Bigg\langle {\frac{|u-u_\text{est}|_2^2}{ |u|_2^2}} \Bigg\rangle_{u,\eta}, \label{lassoloss}\\
    u_\text{est} = \argmin_{x} |G x - y|_2^2 + \alpha |\Psi x|_1 + \beta |x|_2^2.
    \label{genlasso}
\end{gather}
The reconstructed object $u_\text{est}$ is formed by solving a generalized Lasso problem~\cite{10.1214/11-AOS878} with an additional ElasticNet $\ell_2$ term~\cite{10.2307/3647580} (which accelerates convergence as explained below). Here, $\langle~ \rangle_{u,\eta}$ denotes averaging over multiple training objects and noise realizations, and $\Psi$ is a sparsifying transformation under which we expect the object to have a large number of coefficients that are nearly zero. The $\ell_1$ regularization term $|\Psi x|_1=\sum_j |(\Psi x)_j|$ reflects this prior by encouraging sparsity of the vector $\Psi x$~\cite{4472240}. In the imaging settings we treat in this paper, $\Psi$ is the identity operator ($\ell_1$ regularization), but we also develop our sensitivity analysis 
for the 2D/3D anisotropic gradient operator with periodic boundary conditions---total variation (TV) regularization~\cite{RUDIN1992259}---to show the generality of our framework for arbitrary linear transformations $\Psi$. 
We solve \cref{genlasso} using the matrix-free fast iterative soft-thresholding algorithm (FISTA)~\cite{beck2009fast, 7745938}, which allows us to exploit the convolutional structure of the large matrix~$G$~\cite{Antipa:18,lin2021end}.

We work in the paraxial regime, where the response of the physics platform is characterized by a set of point spread functions (PSFs)---the 2D image of a point source on the optical axis---one for each depth and/or wavelength channel~\cite{goodman2005introduction}. The computation of the PSFs consists of spherical wave propagation, followed by a simulation of light scattering at the metasurface, followed by a near-to-far field transformation to compute the electric field in the sensor plane. To model scattering at the metasurface, we use the locally periodic approximation~\cite{Pestourie:18}, in which the large-area metasurface is locally approximated by subwavelength \emph{periodic} unit cells, each containing a nanopillar. We use rigorous coupled-wave analysis~(RCWA)~\cite{liu2012s4} to simulate the complex-valued transmission coefficients of a library of unit cells, from which we construct Chebyshev polynomials that interpolate the transmission coefficients as a function of the nanopillar geometrical parameters~\cite{Pestourie:18}. This differentiable Chebyshev surrogate model allows for rapid simulation of metasurface scattering during optimization.           

From these PSFs, we assemble the measurement matrix $G$. The action of $G$ is to convolve each channel of the object with its corresponding PSF, sum up the resulting contributions to the electric field intensity in the sensor plane, and then crop the field onto the sensor's area. Here, the number of sensor pixels is smaller than the number of pixels in the (potentially multichannel) object, and thus $G$ has fewer rows than columns. As described, we add independent Gaussian noise to the sensor readings, which approximates Poisson shot noise in the limit of large photon flux. 

In our bilevel formulation, the choice of ElasticNet reconstruction ($\beta>0$) instead of pure Lasso ($\beta=0$) serves a crucial purpose \emph{during} the optimization. The initial metasurface geometry leads to a measurement matrix $G$ that is highly unsuitable for compressed sensing; with no $\ell_2$ term, the number of FISTA iterations required for convergence can be orders of magnitude larger for unoptimized metasurface geometries than for optimized geometries. Intuitively, for poorly performing~$G$, nearby pixels are difficult to distinguish, so FISTA spends a large number of iterations making an arbitrary, badly conditioned decision about which nonzero entries to include. With the $\ell_2$ term, the algorithm instead ``gives up'' on this futile task and spreads the values across many entries. In practice, we find that a large initial $\beta$ makes the number of nested iterations roughly \emph{independent} of $G$, fixing the slow-start issue. As we approach the optimized matrix, the need for the $\ell_2$ term diminishes and the optimization chooses to shrink $\beta$ to~$\approx 0$, recovering the pure Lasso (see \cref{fig:2dbeta}). 
\subsection{Sensitivity Analysis of Compressed Sensing Reconstruction}

\label{sec:sens}
For large-scale optimization ($\sim 10^6$ parameters) to be feasible, we must compute the sensitivity of the error $\mathcal{L}$ with respect to the parameters $p,\alpha$, and $\beta$. An adjoint method allows us to backpropagate gradients through our pipeline, at a computational cost similar to that of~$\mathcal{L}$. For the majority of the pipeline, this can be done effortlessly using automatic differentiation tools \cite{innes2019differentiable}. 
\ifdefined\figures
\begin{figure}
    \centering
    \includegraphics[width=0.75\linewidth]{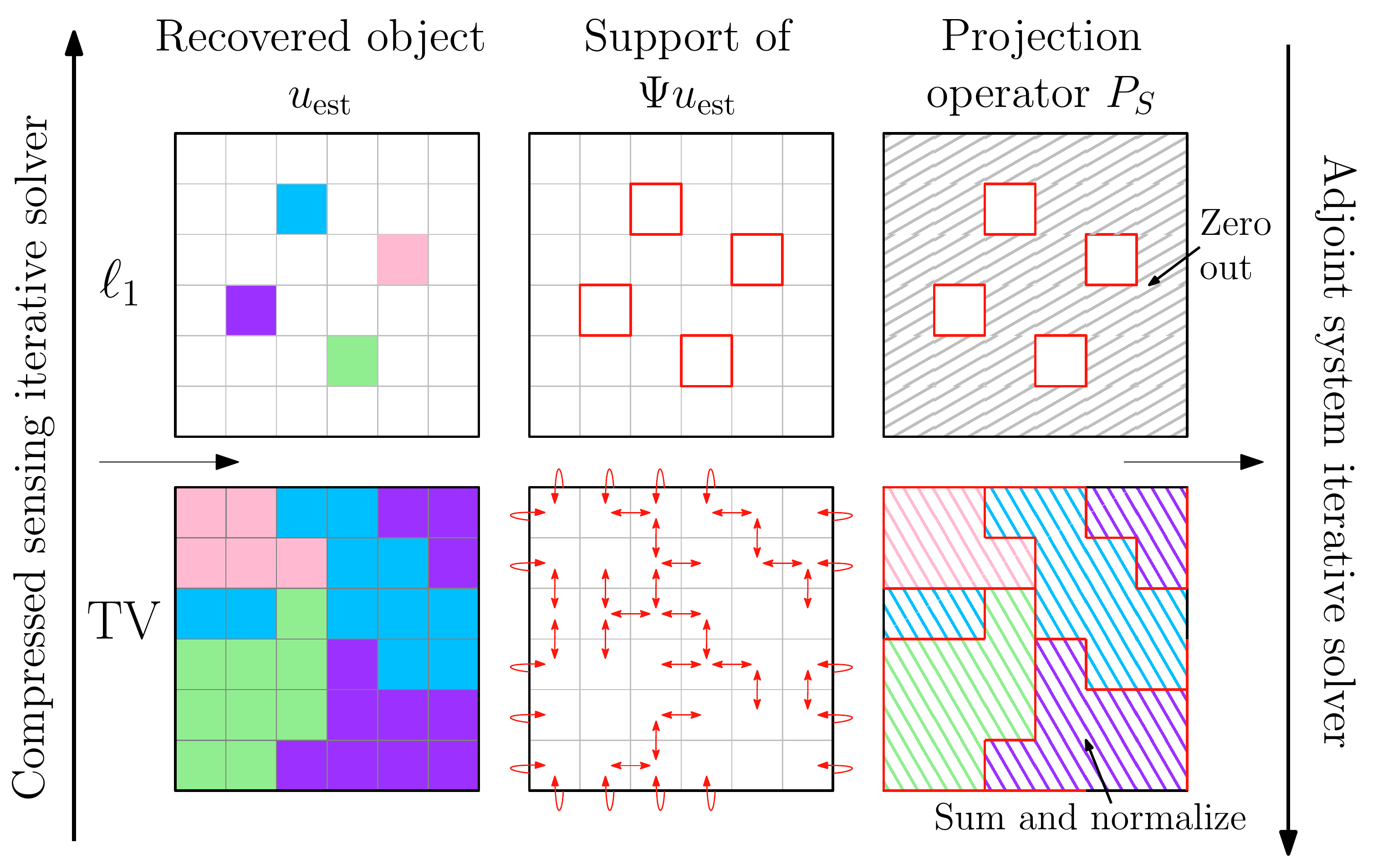}
        \caption{Formation of the projection operator $P_S$, given the Lasso solution found in the forward pass. For simplicity we depict $6 \times 6$ 2D objects, where different colors represent different intensity values, in the cases of both $\ell_1$ and TV regularization. The support of $\Psi u_\text{est}$ is denoted in red (boxes around nonzero pixels for $\ell_1$ regularization, arrows representing nonzero differences for TV regularization). The operator $P_S$ is then used to set up the adjoint system, which is solved in the backward pass.}
    \label{fig:adjoint}
\end{figure}
\else 
\refstepcounter{figure}\label{fig:adjoint}
\fi

However, differentiating through \cref{genlasso} requires special handling. Since we compute the reconstruction $u_\text{est}$ (a vector of length $\approx 10^5$ in our optimizations) through an iterative solver, directly backpropagating through the solver's iterations would require a prohibitive amount of memory. 
Furthermore, due to the nondifferentiability of the $\ell_1$ term $|\Psi u|_1$, the reconstruction $u_\text{est}$ is itself not strictly differentiable with respect to $G$, $\alpha$, and $\beta$.

To surmount these hurdles, our starting point is the KKT conditions for the generalized Lasso \cite{10.1214/12-AOS1003}. We expect that $\Psi u_\text{est}$ is sparse; let $S$ be the support of $\Psi u_\text{est}$ (the set of indices of nonzero values). We can then define $P_S$, the orthogonal projection operator onto the subspace of objects $x$ that satisfy $\support \Psi x \subseteq S$. When $\Psi$ is the identity ($\ell_1$ regularization), $P_S$ restricts an object $x$ to the indices in $S$, zeroing out the other values. When $\Psi$ is the anisotropic gradient operator (TV regularization), $P_S$ sums an object $x$ over each of the piecewise constant regions induced by the sparse gradient $\Psi u_{\text{est}}$, divides each region's sum by the square root of the size of the region (the division ensures $P_S$ is an orthogonal projection, i.e. $P_S = P_S^T$), and sets all elements in that region equal to this summed and normalized value. 
These two cases are illustrated in \cref{fig:adjoint}. The KKT conditions then imply that
\begin{equation}
    \underbrace{(P_S G^T G P_S^T + \beta I)}_{A} \underbrace{u_\text{est}}_{x} = \underbrace{P_S G^Ty - \frac{1}{2} \alpha P_S \Psi^T \sgn{\Psi u_{\text{est}}}}_b,
    \label{fpe}
\end{equation}
which is a square linear system with dimension equal to the rank of $P_S$.  Intuitively, the hard work performed in solving \cref{genlasso} by convex optimization methods is to discover the support $S$. Once $S$ (and the signs $\sgn{\Psi u_{\text{est}}}$ of the elements of $\Psi u_{\text{est}}$) have been found in the forward pass, we know a subgradient of the nondifferentiable $\ell_1$ term $|\Psi u|_1$ at $u_{\text{est}}$. This information allows us to form an efficient backward pass.

Specifically, we can use \cref{fpe} to compute the sensitivity of $u_\text{est}$. Given the gradient $\frac{\partial \mathcal{L}}{\partial u_{est}}$, we write $\frac{\partial \mathcal{L}}{\partial x} = P_S \frac{\partial \mathcal{L}}{\partial u_{\text{est}}}$. Given $\frac{\partial \mathcal{L}}{\partial x}$, we now apply the standard adjoint method for a linear system $Ax = b$~\cite{strang_2019} to find $\frac{\partial \mathcal{L}}{\partial A}$ and $\frac{\partial \mathcal{L}}{\partial b}$ by solving a single linear system with the left operator $A^T$ (the adjoint system). Since $A = P_S G^T G P_S^T$ is symmetric, we can solve this system using the matrix-free conjugate-gradient method~\cite{10.5555/865018}. By backpropagating through the formation of $A$ and $b$ from $G, \alpha$ and $\beta$, we then find the desired sensitivities $\frac{\partial \mathcal{L}}{\partial G}$, $\frac{\partial \mathcal{L}}{\partial \alpha}$, and $\frac{\partial \mathcal{L}}{\partial \beta}$. In this manner, we can backpropagate the gradient through \cref{genlasso} and thus the entire pipeline. An important aspect of our algorithm is that it is ``matrix-free''~\cite{LangvilleMeyer+2011}: in metasurface optics, the measurement matrix is an enormous convolution with a Green's function~\cite{goodman2005introduction} that would be impractical to construct explicitly, but our algorithm is constructed to only employ the matrix implicitly as a linear operator (exploiting FFT-accelerated convolutions~\cite{smith1997scientist}).

We remark that the solution to the generalized Lasso is unique almost surely for the cases of $\ell_1$ and TV regularization \cite{10.1214/13-EJS815, 10.1214/19-EJS1569}, and hence $A$ is nonsingular. In our optimizations, we find the adjoint system to be well-conditioned (it is strictly easier than the convex-optimization forward problem). It is important to note that $\frac{\partial \mathcal{L}}{\partial G}$, $\frac{\partial \mathcal{L}}{\partial \alpha}$, and $\frac{\partial \mathcal{L}}{\partial \beta}$ are only \emph{subgradients}, as the support of $\Psi u_\text{est}$ can change. However, the support of $\Psi u_\text{est}$ is locally constant almost everywhere~\cite{10.1214/12-AOS1003}; in practice, we observe excellent results when plugging the subgradient $\frac{\partial \mathcal{L}}{\partial p}$ into standard gradient-based optimization algorithms.  We employ the stochastic-gradient algorithm Adam \cite{kingma2014adam} because our training objects~$u$ are drawn from a random distribution of sparse objects.
 
\section{Results}

\subsection{Single Channel Imaging with Small Sensor}

\ifdefined\figures
\begin{figure}
        \centering
    \begin{subfigure}{\textwidth} 
 \refstepcounter{subfigure}\label{fig:2dpsf}
        \refstepcounter{subfigure}\label{fig:2dobj}
        \refstepcounter{subfigure}\label{fig:2dimg}
        \refstepcounter{subfigure}\label{fig:2drec}
        \refstepcounter{subfigure}\label{fig:2dphase}
        \refstepcounter{subfigure}\label{fig:2dalpha}
        \refstepcounter{subfigure}\label{fig:2dbeta}
        \refstepcounter{subfigure}\label{fig:2drmse}
        \refstepcounter{subfigure}\label{fig:2dmeta}
    \end{subfigure}\includegraphics[width=0.7\textwidth]{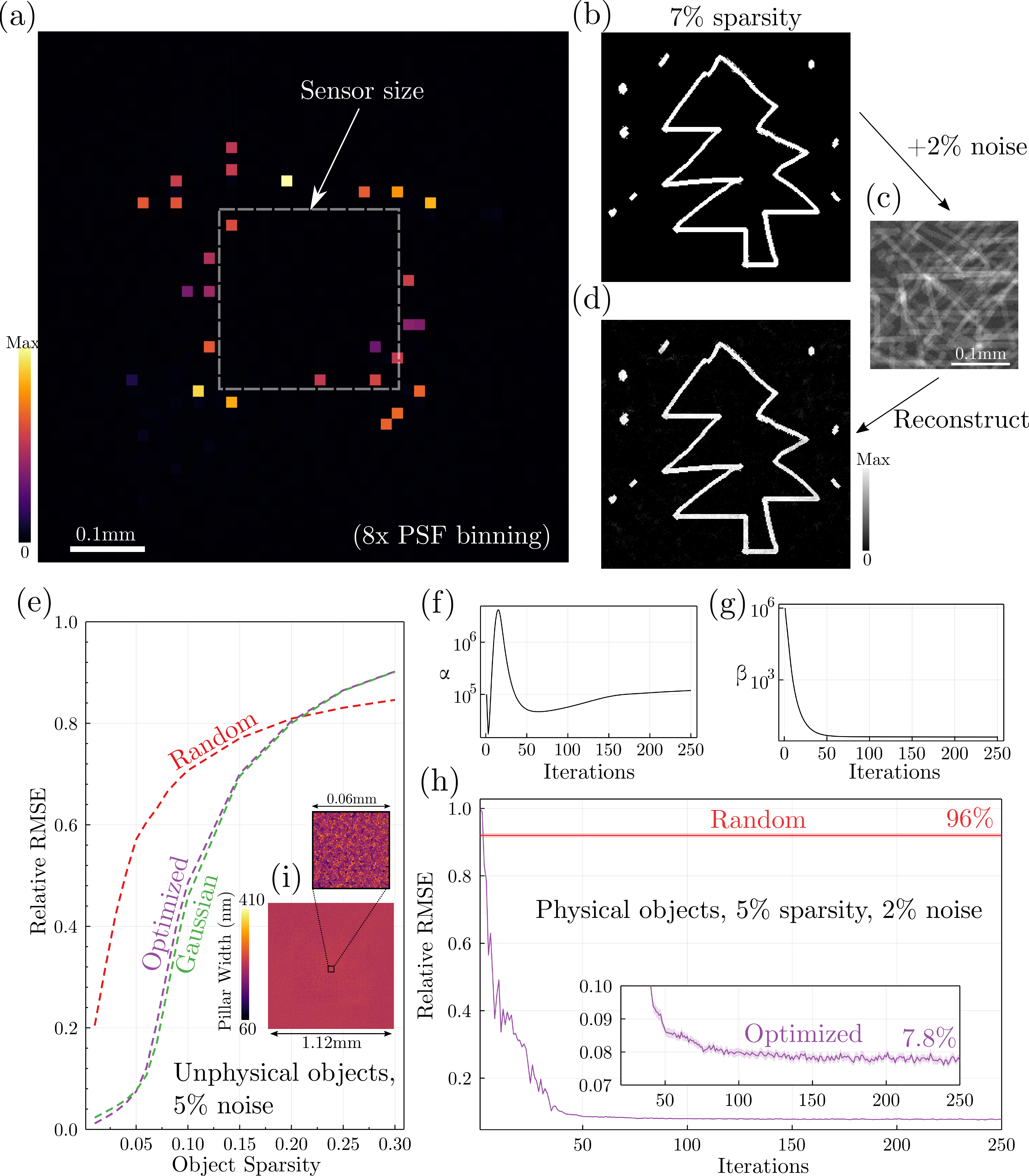}
    \caption{Metasurface 2D imager with compressed sensing. \textbf{(a)} Optimized point spread function, a multifocal metalens with 27 foci ($8\times$ binned to emphasize lens positions). 
    \textbf{(b)} Ground truth $256 \times 256$ sparse object (7\% sparsity) with wavelength $470\mathrm{nm}$. 
    \textbf{(c)} Image formed from ground truth object by the optimized design, with additive Gaussian noise (2\%). 
    \textbf{(d)} CS reconstruction of the ground truth object from the noisy image.  
    \textbf{(e)} Reconstruction error on \emph{unphysical objects} whose non-zero values are drawn from a distribution with mean 0 as a function of object sparsity, for the optimized design, a random design, and the ideal Gaussian baseline under the same noise setting (5\%). 
    \textbf{(f)} Evolution of $\alpha$ during optimization. 
    \textbf{(g)} Evolution of $\beta$ during optimization. 
    \textbf{(h)} Convergence of the reconstruction error during training on random objects with $5\%$ sparsity. The objects have 3250 positive values uniformly sampled from the volume, with intensity values uniformly drawn from the interval $[0.8,1.2]$. We also plot the error of a random metasurface. 
    \textbf{(i)} The $1.12\mathrm{mm} \times 1.12\mathrm{mm}$ TiO$_2$ metasurface (refractive index $n \approx 2.4$) on a silica substrate, consisting of a grid of $2048 \times 2048$ subwavelength square unit cells, where each unit cell has width $\approx 470 \mathrm{nm}$. The square pillar in each cell has a height of $600\mathrm{nm}$, and a width ranging from $\approx 60\mathrm{nm}$ to $\approx 410\mathrm{nm}$. The metasurface is depicted as a heatmap of the pillar widths, and we also zoom in on a $0.06 \times 0.06\mathrm{mm}$ section of the metasurface.}
    \label{fig:2d}
\end{figure}
\else 
\refstepcounter{figure}\label{fig:2d}
 \refstepcounter{subfigure}\label{fig:2dpsf}
        \refstepcounter{subfigure}\label{fig:2dobj}
        \refstepcounter{subfigure}\label{fig:2dimg}
        \refstepcounter{subfigure}\label{fig:2drec}
        \refstepcounter{subfigure}\label{fig:2dphase}
        \refstepcounter{subfigure}\label{fig:2dalpha}
        \refstepcounter{subfigure}\label{fig:2dbeta}
        \refstepcounter{subfigure}\label{fig:2drmse}
        \refstepcounter{subfigure}\label{fig:2dmeta}
\fi
 \label{sec:2d}
We first apply our optimization technique to single-channel imaging of sparse objects in the visible regime. We consider the case of a \emph{small sensor} that has fewer pixels ($128 \times 128$) than the object we wish to reconstruct ($256 \times 256$). This is an underdetermined situation in which a traditional lens would perform very poorly because the object would not fit on the sensor. However, working under the assumption that our object is sparse in the spatial domain (thus we set $\Psi$ to the identity), one may hope for a ``multiplexing'' design that compresses the object onto the small sensor in such a way that the image can be deconvolved effectively through CS. It is unclear \emph{a~priori} what metasurface geometry would best realize this goal. 

As shown in \cref{fig:2d}, the end-to-end optimization \emph{spontaneously configures} an arrangement of lens-like focusing elements, leading to a number of sharp peaks in the far field (\cref{fig:2dpsf}). Over 250 iterations of optimization on training objects with 5\% sparsity (the proportion of nonzero values), the metasurface---initialized as an array of identical nanopillars---evolves into the structure shown in \cref{fig:2dmeta}. The relative RMSE (the square root of the relative MSE $\mathcal{L}$) converges to $7.8\%$ (\cref{fig:2drmse}). This is an order of magnitude improvement over a randomly structured metasurface (96\% relative RMSE). 

Effectively, the optimized metasurface system takes many copies of a sparse object (\cref{fig:2dobj} shows a test object with 7\% sparsity) and ``interlaces'' them on the smaller sensor (\cref{fig:2dimg}). The object is faithfully recovered (18\% relative RMSE, \cref{fig:2drec}) even though the small sensor image is further corrupted by noise (2\%). Our optimization also automates the choice of the reconstruction-algorithm hyperparameters $\alpha$ and $\beta$. In particular, $\alpha$ varies significantly on a logarithmic scale (\cref{fig:2dalpha}) and settles at approximately $10^5$, while $\beta$ shrinks to a negligible value that is orders of magnitude smaller than $\alpha$ (\cref{fig:2dbeta}).

We now discuss how the measurement matrix of our optimized system compares to the ``gold standard'' of compressed sensing, a Gaussian random matrix. Such a matrix is \emph{unphysical} for our optical system, but by comparing to it we can characterize the performance of our system relative to \emph{all} possible matrices, whether physical or not. Representing Gaussian matrices would require a prohibitive amount of memory, so we use partial Gaussian circulant matrices as a computationally tractable equivalent~\cite{1660793}. As shown in \cref{fig:2dphase}, a random metasurface significantly underperforms a Gaussian matrix. Remarkably, however, the optimized metasurface matches the performance of the Gaussian matrix throughout the entire error-sparsity ``phase transition''~\cite{blanchard2013toward} (the upswing segments of the error-sparsity relations in \cref{fig:2dphase}). 

We qualify this apparent optimality result by noting that the comparison in \cref{fig:2dphase} is made with \emph{unphysical} objects. (In the following, ``unphysical'' objects have mean zero and can have negative values; ``physical'' objects have only non-negative values.) On physical objects, optical systems are less robust to noise. This is due to the non-negativity of the measurement matrix of the metasurface system, which is a fundamental optical limitation. Intuitively, in the case of physical objects and measurement matrices, light can only be added to the sensor and not subtracted, making it harder to form sharp noise-robust features. This limitation is not faced by the Gaussian measurement matrices used as benchmarks in \cref{fig:2dphase}, which have negative values and are therefore unphysical (not realizable with optics). 

Mathematically, recalling that $\sigma \propto |Gu|_1$ in our noise model from \cref{eq:noisemodel}, 
the worse performance on physical objects manifests as a relative increase in the mean intensity magnitude of the image relative to unphysical objects (which include negative ``intensities''), quantified by the following ratio: 
\begin{equation}
    \text{image mean gap of }G = \frac{\langle |G u|_1 \rangle_{\text{phys.}}}{\langle |G u|_1 \rangle_{\text{unphys.}}} \cdot \frac{\langle |X u|_1 \rangle_{\text{unphys.}}}{\langle |X u|_1 \rangle_{\text{phys.}}},
    \label{image_mean_gap}
\end{equation}
where $G$ is the measurement matrix of the metasurface system, $X$ is a Gaussian matrix, and $\langle \cdot \rangle$ denotes averaging over particular distributions of physical and unphysical objects. The ``image mean gap'' characterizes the ability of the metasurface system to form sharp noise-robust image features on \emph{physical} non-negative objects. From another perspective, the image mean gap measures the ``noisiness'' of the PSF, i.e. its deviation from a sharply focused lens~$G=I$.

We can use the image mean gap to link the results of \cref{fig:2dphase} to the physical case. We consider our training object distribution, where objects have $5\%$ sparsity and their nonzero values are drawn uniformly from the interval $[0.8,1.2]$. In this setting, the image mean gap is $\approx 5.4$ for a random metasurface and $\approx 2.1$ for the optimized structure. This poorer image mean gap, combined with the random metasurface's poorer performance on unphysical objects in \cref{fig:2dphase}, explains the random metasurface's very poor performance on physical objects (96\% relative RMSE). 
The image mean gap of $\approx 2.1$ ($2\times$ improvement) for the optimized structure is significantly better: it implies that our metasurface system performs equivalently to an \emph{unphysical} Gaussian matrix system that has $\approx 2.1\times$ more noise. It is unclear if the image mean gap can be reduced even further while maintaining the design's optimality on unphysical objects. One might hope that a reduced number of foci in the design, leading to less overlap of the object ``copies'' on the sensor, would lead to sharper image features on non-negative objects and thus a better image mean gap. But this could come at the expense of optimality on unphysical objects due to reduced multiplexing of the object on the sensor, illustrating a trade-off between the two factors. We leave more refined lower bound analyses for future work. 

For physical imaging situations, we emphasize the importance of sparsity \emph{of the measurement matrix}, which corresponds to lens-like PSFs in the convolutional model. 
Indeed, under a non-negativity constraint, one can show that \emph{sparse} random matrices can perform well~\cite{indyk}, and furthermore that convolution with certain random sparse patterns can lead to excellent performance~\cite{Marcos:16}. Without any human input, the end-to-end optimization forms a \emph{sparse} PSF that fits this characterization with optimal choices of the number and positions of lenses. In doing so, the optimization forms a measurement matrix with a reduced image mean gap ($\approx 2.1$) and a near-exact match with a Gaussian matrix on unphysical objects, therefore finding a \emph{physical} metasurface geometry that significantly improves over random scattering patterns.

\subsection{High Resolution 3D Imaging with PCM}

\label{sec:gss}
\ifdefined\figures
\begin{figure}
    \makebox[\textwidth][c]{\includegraphics[width=\textwidth, trim={0cm 0.0cm 0cm 0cm}]{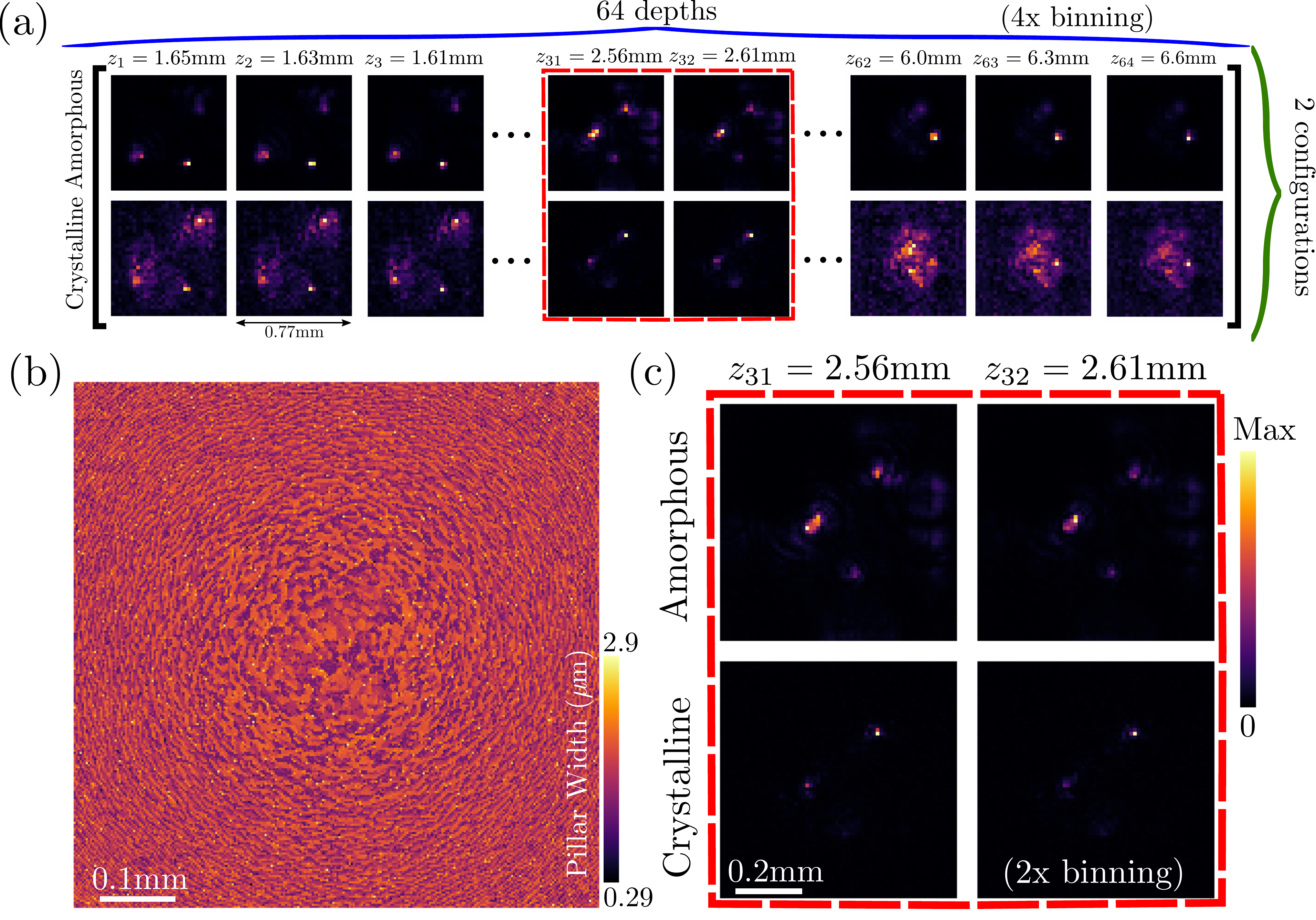}}
    \begin{subfigure}{\textwidth} 
    \refstepcounter{subfigure}\label{fig:3dl1psf} 
        \refstepcounter{subfigure}\label{fig:3dl1meta}
        \refstepcounter{subfigure}\label{fig:3dl1disperse}
    \end{subfigure}\caption{Design of GSS4T1 3D imager with compressed sensing. \textbf{(a)} Optimized point spread functions, which exhibit high-frequency features, depth sensitivity, and phase change contrast (4x binned to emphasize lens positions). \textbf{(b)} The $0.77\mathrm{mm} \times 0.77\mathrm{mm}$ GSS4T1 metasurface on a CaF$_2$ substrate, consisting of a grid of $256 \times 256$ subwavelength square unit cells,. Each unit cell has width $\approx 3 \mathrm{\mu m}$ and index phase contrast $\Delta n = 1.40$. The square pillar in each cell has a height of $1.1\mu\mathrm{m}$, and width ranging from $\approx 0.29\mu\mathrm{m}$ to $\approx 2.9\mu\mathrm{    m}$. The metasurface is depicted as a heatmap of the pillar widths. \textbf{(c)} Zoom-in of PSFs at adjacent depths, illustrating the depth sensitivity and phase change contrast. }
    \label{fig:gss}
\end{figure}
\else 
\refstepcounter{figure}\label{fig:gss}
    \refstepcounter{subfigure}\label{fig:3dl1psf} 
        \refstepcounter{subfigure}\label{fig:3dl1meta}
        \refstepcounter{subfigure}\label{fig:3dl1disperse}
\fi

To show the flexibility of our framework, we now consider a multichannel 64-depth imaging problem in the mid-infrared regime. In particular, we endeavor to image $\ell_1$-sparse, volumetric distributions of point sources at very fine depth resolutions. This leads to a severely underdetermined problem where an object with discretization $32 \times 32 \times 64\textrm{ depths}$ is captured on a $96 \times 96$ sensor. To alleviate this problem, we incorporate a \emph{reconfigurable} material (PCM), GSS4T1, which can undergo a rapid phase change from amorphous to crystalline states via electrical switching~\cite{zhang2019broadband}. Thus, a GSS4T1 metasurface allows us to obtain two images from a single structure design with two material phases of refractive index contrast $\Delta n = 1.40$ (\cref{fig:3dgsspillar}). This presents a fascinating problem for end-to-end optimization: the fineness of the required depth resolution necessitates high depth sensitivity, and the metasurface design must also ensure contrast between the images corresponding to the two material phases (but the same geometry) so that the second shot provides useful additional information. Our two-shot scheme using a PCM is significantly faster than multi-shot schemes that use digital micromirror devices or spatial light modulators~\cite{willett2011compressed}. 

In this setting, the optimization produces the metasurface shown in \cref{fig:3dl1meta}. The PSFs of the optimized metasurface exhibit high-frequency lens-like features (\cref{fig:3dl1psf}), necessary for good lateral resolution. Moreover, the PSFs demonstrate spatial dispersion and phase change contrast (\cref{fig:3dl1disperse}) despite significant physical challenges. For instance, the thickness of the GSS4T1 metasurface is a tenth of the wavelength in the mid-infrared regime, limiting the amount of wave scattering and dispersion available at the interface. Over the depth range, we notice dramatic evolution in the point-spread function, as various lenses of the multifocal design come in and out of focus. However, physical limitations prevent the design from achieving such dramatic shifts between every pair of adjacent depths. Instead, the optimization tends to judiciously shift lens positions between adjacent depths by a few pixels, thereby shifting the images of adjacent channels on the sensor sufficiently to achieve the desired $z$-resolution. Crucially, the optimized design achieves these goals for the images of both configurations of the GSS4T1 metasurface. Furthermore, it achieves high contrast between configurations for the same depth channel, thus collecting \emph{complementary} information from the two shots.

\ifdefined\figures
\begin{figure}
    \makebox[\textwidth][c]{\includegraphics[width=\textwidth, trim={0 0cm 0 0}]{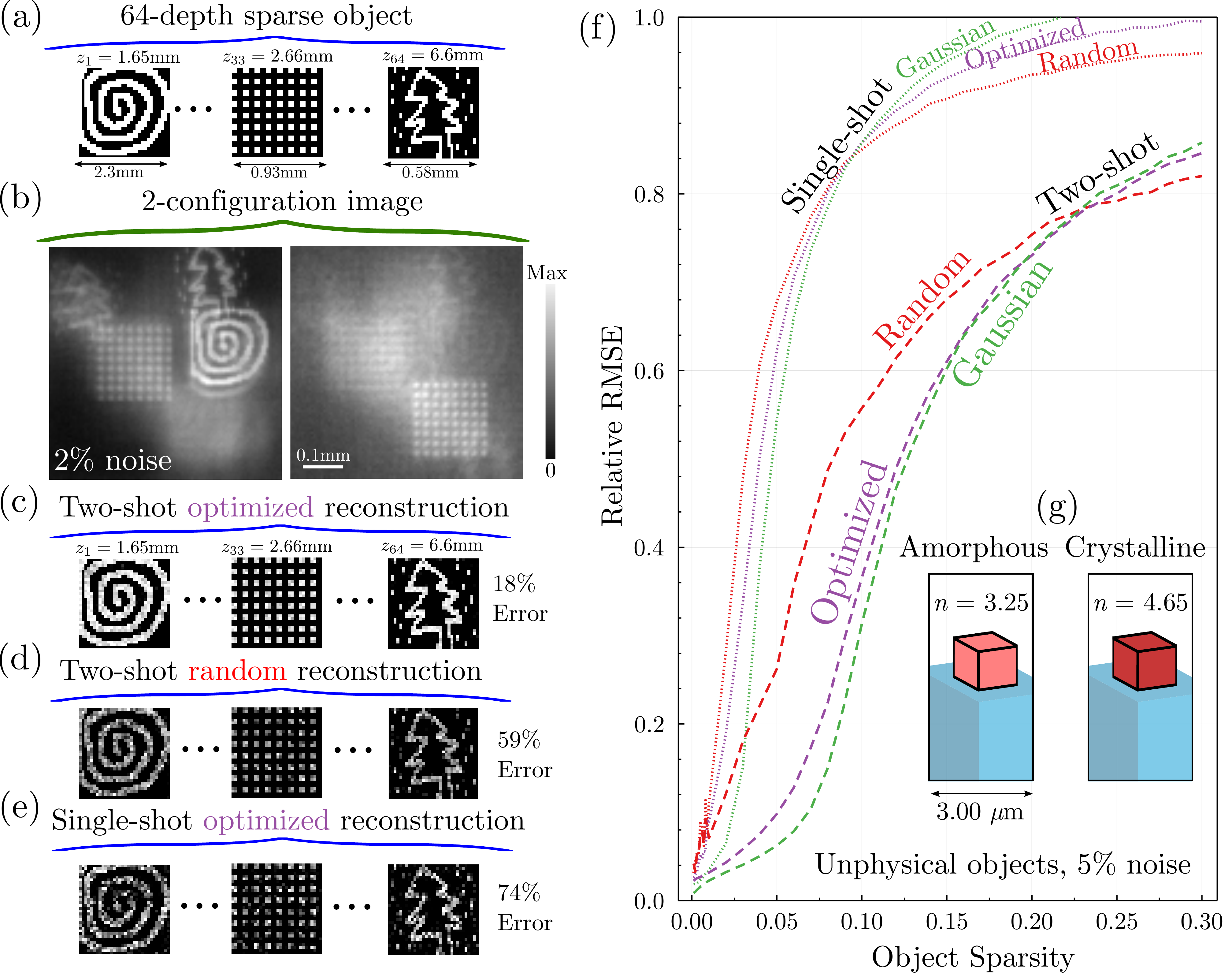}}
    \begin{subfigure}{\textwidth} 
    \refstepcounter{subfigure}\label{fig:3dl1obj}
        \refstepcounter{subfigure}\label{fig:3dl1img}
        \refstepcounter{subfigure}\label{fig:3dl1rec}
        \refstepcounter{subfigure}\label{fig:3dl1random}
        \refstepcounter{subfigure}\label{fig:3dl1single}
        \refstepcounter{subfigure}\label{fig:3dgsspillar}
        \refstepcounter{subfigure}\label{fig:3dl1phase}
    \end{subfigure}\caption{Analysis of GSS4T1 3D imager with compressed sensing. 
        \textbf{(a)} A 64-depth sparse ground truth test object ($\approx 1.5\%$ sparsity, or 1000 non-zero values) with wavelength $3.2 \mu\mathrm{m}$, and the depth channels inversely spaced from $1.65\mathrm{mm}$ to $6.6\mathrm{mm}$. Patterns of varying structure are ``hidden'' in three of the 64 channels, plotted in grayscale. \textbf{(b)} Two images formed of the ground truth object by the amorphous and crystalline metasurface states of the optimized two-shot structure, with additive Gaussian noise (1\%). 
        \textbf{(c)} CS reconstruction of the ground truth object for the optimized two-shot system (18\% relative RMSE). 
        \textbf{(d)} CS reconstruction of the ground truth object for a two-shot system with a random metasurface (59\% relative RMSE).  \textbf{(e)} CS reconstruction of the ground truth object for an optimized single-shot system (74\% relative RMSE). 
        \textbf{(f)} Reconstruction error on \emph{unphysical objects} with mean 0 as a function of object sparsity for the optimized design, a random design, and an ideal Gaussian baseline, and the corresponding single-shot designs under the same noise setting (5\%). (The single-shot end-to-end design is directly optimized for the single shot case, and the single-shot Gaussian baseline has half as many rows as the two-shot matrix.)
        \textbf{(g)} GSS4T1 metasurface pillar in amorphous ($n = 3.25$) and crystalline ($n = 4.65$) states.}
    \label{fig:gss2}
\end{figure} 
\else 
\refstepcounter{figure}\label{fig:gss2}
    \refstepcounter{subfigure}\label{fig:3dl1obj}
        \refstepcounter{subfigure}\label{fig:3dl1img}
        \refstepcounter{subfigure}\label{fig:3dl1rec}
        \refstepcounter{subfigure}\label{fig:3dl1random}
        \refstepcounter{subfigure}\label{fig:3dl1single}
        \refstepcounter{subfigure}\label{fig:3dgsspillar}
        \refstepcounter{subfigure}\label{fig:3dl1phase}
\fi
While these qualitative considerations give us a clue of the 3D imaging system's resolving power, we now analyze the system's CS performance on complex sparse scenes. In \cref{fig:3dl1phase}, we plot the reconstruction error as a function of object sparsity on \emph{unphysical objects} for the end-to-end optimized design, a random metasurface, and a partial Gaussian circulant baseline. With 5\% noise, we observe that the optimized structure closely (but not perfectly) matches the performance of an ideal Gaussian matrix, while substantially outperforming random metasurfaces (e.g. $\approx 3\times$ smaller RMSE at 3\% sparsity). Now, 
we study the performance gap due to the non-negativity of the imaging system. At 3\% object sparsity, the image mean gap (\cref{image_mean_gap}) between unphysical objects and non-negative training objects with intensity values in the range $[0.8,1.2]$ is 46 for a random structure and 18 for an optimized structure. Once again, the image mean gap is worse for a random structure ($\times 2.5$), showing how the end-to-end optimization (which is trained on physical non-negative objects) favors PSFs that are more lens-like and lead to sharper image features. The lens-like features of the optimized PSFs (\cref{fig:3dl1psf,fig:3dl1disperse}) are not perfectly focused, however. This is reflected in the image mean gap of 18 for the optimized structure, which is still relatively large; incorporating more complex nanophotonic structures than a single-layer metasurface may alleviate this issue.

We also perform a visual test of the system using a physical ground truth object with $\approx 1000$ nonzero values ($\approx 1.5\%$ sparsity) where patterns of varying structure have been hidden at 3 of the 64 depths. This ground truth object differs substantially from the objects in the training distribution which have uniformly distributed nonzero values across all depths. Nevertheless, the optimized imaging system accurately recovers the support and forms a reconstruction with $18\%$ relative RMSE (\cref{fig:3dl1rec}), despite the image being corrupted with $2\%$ noise, showing how the system generalizes to out-of-distribution sparse data.
In comparison to the optimized system, a random metasurface system achieves a much less accurate reconstruction of the same object (59\% relative RMSE, \cref{fig:3dl1random}), and fails to accurately recover the support. 
We also note the significant improvement over \emph{single-shot} designs where the GSS4T1 metasurface is kept in only its crystalline state. On the same ground truth object with $3\%$ sparsity (\cref{fig:3dl1obj}), an end-to-end optimized single-shot design obtains a significantly worse reconstruction (74\% relative RMSE, \cref{fig:3dl1single}). This is due in large part to the delayed sparsity-error phase transition for the two-shot design (the rightward shift and slower incline of the sparsity-error relations in \cref{fig:3dl1phase}), emphasizing the crucial role of the material index contrast and the ability of our end-to-end optimization to exploit it. 
\section{Conclusions and Outlook}

\label{sec:conclusion}

Our results show how end-to-end optimization can navigate complicated, physically-constrained optimization spaces to discover measurement matrices with improved compressed-sensing performance. In certain cases where there are enough physical degrees of freedom for a given task (such as 2D imaging with a small sensor), the optimization discovers a measurement matrix that approaches the performance of a Gaussian matrix for \emph{noise-robust} compressed sensing. Even under severe physical constraints as in the case of a deeply sub-wavelength PCM interface, end-to-end optimization can balance a complicated set of considerations to obtain measurement matrices which significantly outperform random two-shot designs and optimized single-shot designs for high-resolution 3D imaging.  We expect that our technique---which, in generality, allows one to pair any measurement system with an $\ell_1$-regularized CS minimization algorithm and optimize the system \emph{end-to-end}---will find use in a wide range of CS recovery systems whose design involves a large number of degrees of freedom.

In the design of such systems, it is crucial to take into account the type of prior imposed on the object. We note that our end-to-end framework supports any linear sparsifying transformation $\Psi$; we need only be equipped with the projection operator $P_S$ to perform the sensitivity analysis. Future work may build upon our method by performing end-to-end optimization of physical systems that incorporate more complicated non-linear priors that cannot be straightforwardly put in the form of the generalized Lasso. 

{
Several underdetermined imaging applications may also warrant modifications to the end-to-end framework presented here. For instance, one may wish to encourage robustness across a \emph{continuous range} of wavelengths or depths, as opposed to fixing channels at a particular wavelength or depth. A possible application of such an objective would be broadband imaging; in the supplement, we include a study of our existing single-channel imaging system's broadband performance \emph{without} such an objective (Supplementary Fig.~1).} 
{
It would also be informative to further investigate how the noise level and noise type influence the optimized design. Indeed, for task-specific imaging objectives (as opposed to generic object reconstruction), prior work has found that noise levels can strongly influence the optimized design~\cite{doi:10.34133/2022/9825738}. For our setting of additive Gaussian noise with an $\ell_1$ sparsity prior, we find that the optimized designs predictably generalize to different noise levels (Supplementary Fig.~2); but such results for task-specific imaging~\cite{doi:10.34133/2022/9825738} suggest that this would no longer be the case with more complicated priors and/or different noise models. It would be useful to investigate this in future work. Finally, while there is no clear substitute for the end-to-end CS objective in terms of tractability and generality, it would be interesting to investigate incorporating heuristic objectives to speed up the optimization in its early stages. One candidate would be the local condition number metric proposed in previous work \cite{Antipa:18}.
}

{More broadly}, as one leverages richer physics to extract more and more information from a scene with limited sensor data, we anticipate a growing need to cope with severely underdetermined problems. Our end-to-end framework, which supports arbitrary multichannel (and multiconfiguration) settings, paves the way for underdetermined nanophotonics imaging systems that can extract hyperfine depth, spectral, and/or polarization information from a single shot. The ultracompact all-dielectric metasurface designs we have presented are directly amenable to large-scale fabrication platforms~\cite{li2021inverse} and may enable the realization of metasurface cameras for imaging sparse scenes (e.g., locating stars in the night sky).  We are particularly excited about future applications of our technique to thermal imaging, where one may endeavor to construct a full spectral imager that can reconstruct both emissivity and temperature profiles from a single shot. Furthermore, by incorporating multi-layered metasurfaces and photonic crystal slabs, one may exploit even richer \emph{non-local} physics and develop imaging systems whose measurement matrix differs substantially in structure from a simple convolutional model. In this case, there are few conventional principles to guide the design; it is our hope that end-to-end optimization with compressed sensing will allow one to exploit the full power of nanophotonics in challenging underdetermined imaging problems. \newline

\noindent\textbf{Research Funding:} This work was supported in part by the Simons Foundation and by the U.S. Army Research Office through the Institute for Soldier Nanotechnologies under Award No. W911NF-18-2-0048.

\noindent\textbf{Conflict of interest statement:} The authors declare no conflicts of interest regarding this article.

\noindent\textbf{Data availability:} Data underlying the results presented in this paper may be obtained from the authors upon
reasonable request. The code used is publicly available at \url{https://github.com/gaurav-arya/ImagingOpt.jl}. 

\noindent\textbf{Supplementary information:} In the supplement, two additional figures, Supplementary Fig.~1 and Supplementary Fig.~2, are provided, respectively describing the sensitivity of the optimized systems to wavelength aberrations, and the response of the optimized systems to varying noise levels. In addition, an animation of the optimization process for the design depicted in \cref{fig:2d} is provided.

\bibliography{bib}

\appendix

\renewcommand{\figurename}{Supplementary Fig.}
\setcounter{figure}{0}

{
\section{Comparing optimized system with wavelength aberrations to random metasurface systems}

In this analysis, we evaluate the robustness of our design's imaging performance under broadband illumination. Our imaging systems in the main text were optimized for objects emitting at a fixed wavelength $\lambda$, achieving significant performance improvements over a random metasurface structure. This raises the question: does this performance improvement persist even when we introduce aberrations into the point-spread function by using light at a different wavelength $\lambda + \Delta \lambda$?

\ifdefined\figures
\begin{figure}
        \centering
    \begin{subfigure}{\textwidth} 
        \refstepcounter{subfigure}\label{fig:2dwavelength_psf}
        \refstepcounter{subfigure}\label{fig:2dwavelength_img}
        \refstepcounter{subfigure}\label{fig:2dwavelength_rec}
    \end{subfigure}\includegraphics[width=0.8\textwidth]{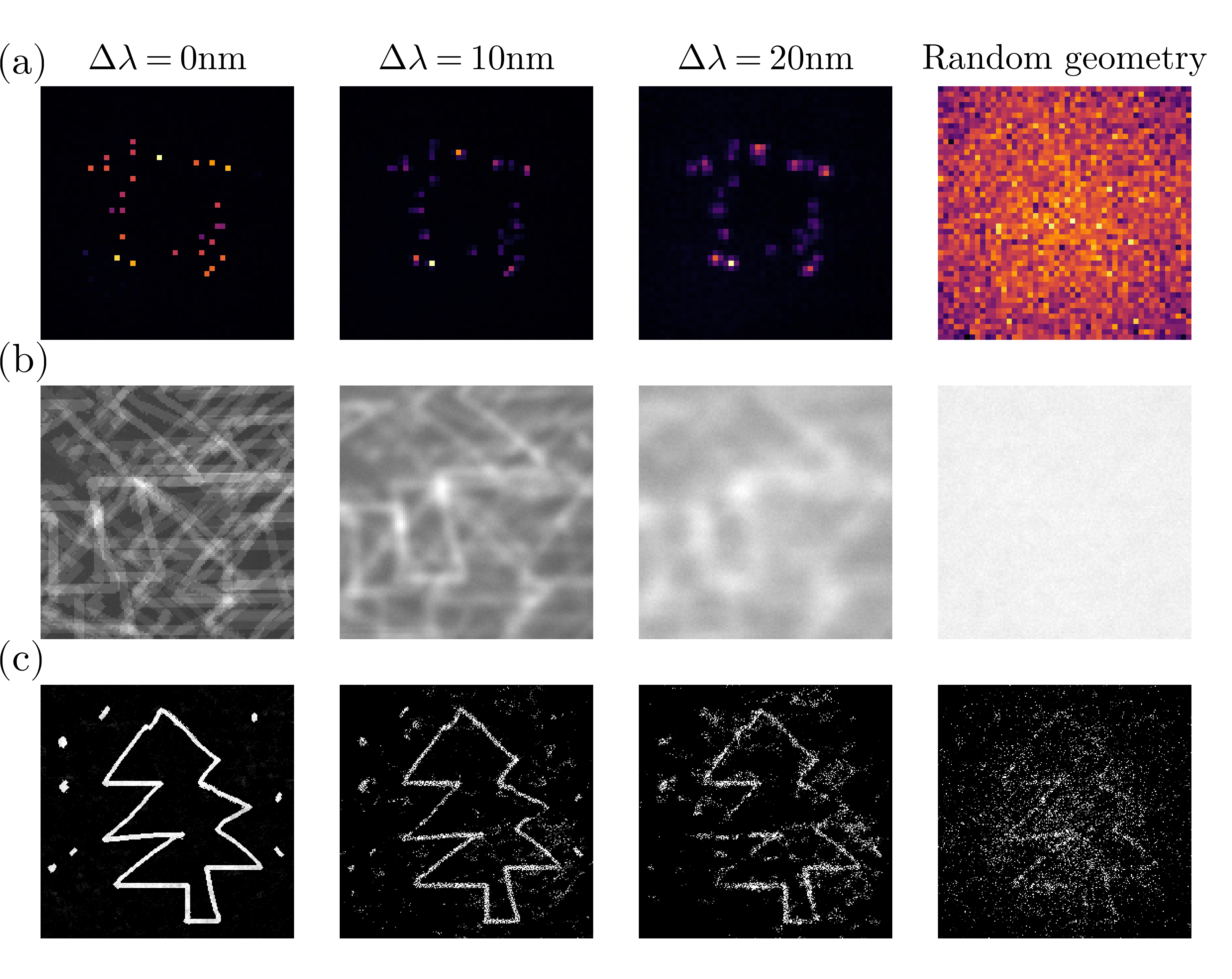}
    \caption{{Wavelength dependence of metasurface 2D imager with compressed sensing. \textbf{(a)} PSFs for optimized system at wavelength $\lambda + \Delta \lambda$ for $\lambda = 470\mathrm{nm}$ and $\Delta \lambda \in \{0\mathrm{nm}, 10\mathrm{nm}, 50\mathrm{nm}\}$, and for a random metasurface. All PSFs are 8x binned as in \cref{fig:2dpsf}. \textbf{(b)} Images formed using ground truth object from \cref{fig:2dobj} for each PSF. \textbf{(c)} CS reconstruction of ground truth object for each PSF.}}
    \label{fig:2dwavelength}
\end{figure}
\else 
\refstepcounter{figure}\label{fig:2d_wavelength}
\fi

Supplementary Fig.~\ref{fig:2dwavelength_psf} shows the PSFs of the single-channel underdetermined imaging case (\cref{fig:2d}) as the wavelength $\lambda = 470\mathrm{nm}$ is perturbed by $\Delta \lambda = 10\mathrm{nm}$ and $\Delta \lambda = 50\mathrm{nm}$. It also shows the PSF of a random structure. We evaluate each of these designs on our original test object (\cref{fig:2dobj}, 7\% sparsity); for each design, we retune the $\ell_1$ regularization parameter $\alpha$ to ensure a fair comparison.
We find that the resultant images (Supplementary Fig.~\ref{fig:2dwavelength_img}) become blurrier with higher aberration due to defocusing of the lenses, resulting in significantly poorer reconstruction performance (Supplementary Fig.~\ref{fig:2dwavelength_rec}). However, the essential PSF structure persists, and we still observe a noticeable improvements over random structures for $\Delta \lambda \in \{10\mathrm{nm}, 50\mathrm{nm}\}$, meaning that the benefits of end-to-end design can be robust to significant aberrations relative to the computational model.

We note that our end-to-end setup did not explicitly include any objectives for encouraging broadband performance. In future work, for applications demanding stable reconstruction performance across a wide range of wavelengths such as broadband thermal imaging, it would be fruitful to explicitly optimize for robustness across a broadband range.

\section{Noise sensitivity}

In this analysis, we evaluate the performance of our designs under different noise levels.
In our end-to-end designs, we ran our optimization at a specific noise level. Supplementary Fig.~\ref{fig:2dnoise} shows the RMSE of our single-channel imaging system (\cref{fig:2d}) against the noise level. A fixed test object (the same as depicted in \cref{fig:2dobj}) is used, and $\alpha$ is retuned for each noise level. We observe that the RMSE follows a roughly linear relationship with noise for small noise levels. 
\ifdefined\figures 
\begin{figure}
        \centering
    \includegraphics[width=0.6\textwidth]{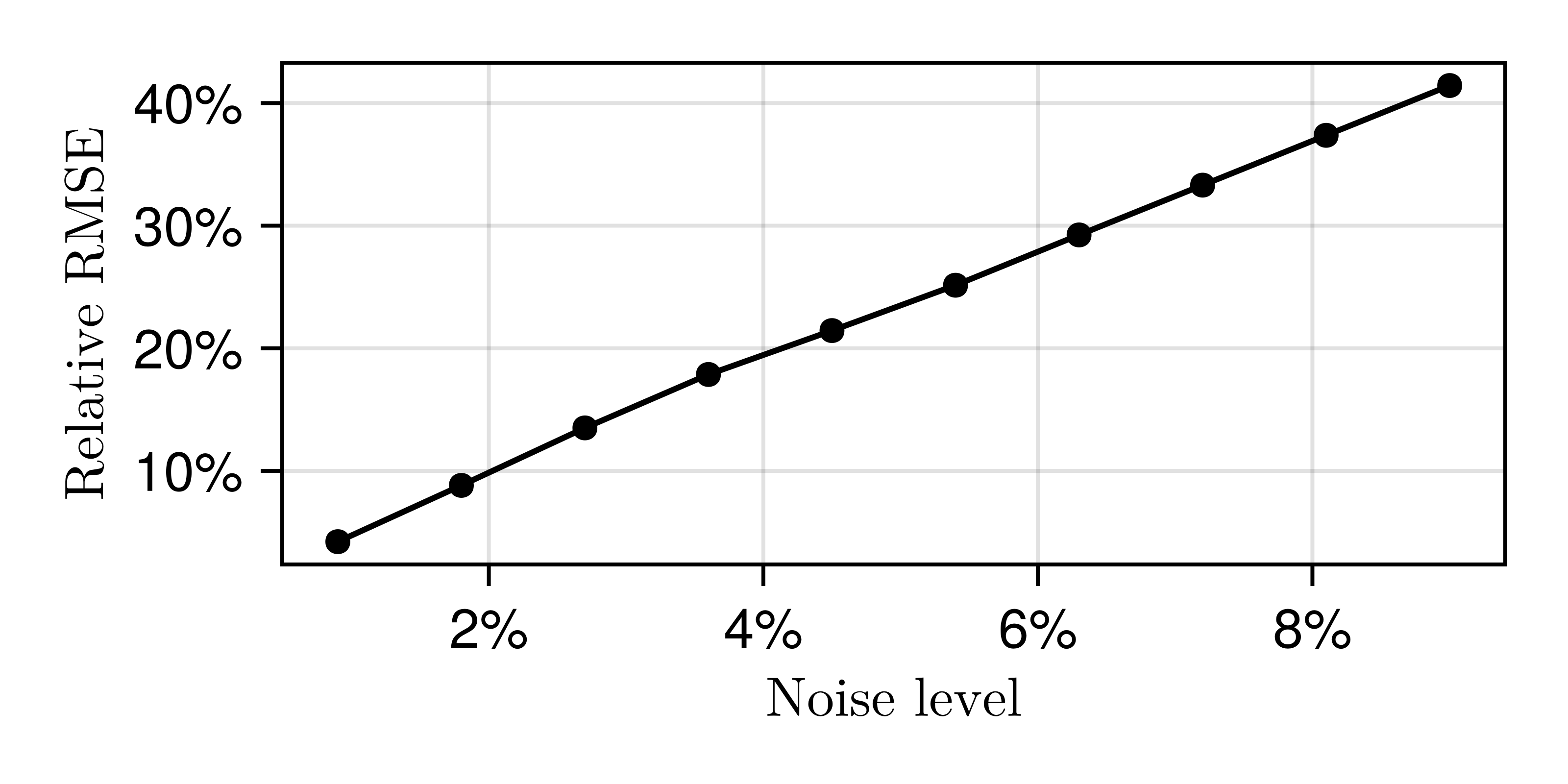}
    \caption{{Relative RMSE versus noise level for optimized metasurface 2D imager depicted in \cref{fig:2d}, using ground truth test object from \cref{fig:2dobj}.}}
    \label{fig:2dnoise}
\end{figure}
\else 
\refstepcounter{figure}\label{fig:2dnoise}
\fi

}

\end{document}